\documentstyle[12pt,cite,epsfig]{article}

\begin{document}

\renewcommand{\thefootnote}{\fnsymbol{footnote}}

\begin{flushright}
 {\small
   SLAC--PUB--8772\\
   February 2001\\}
\end{flushright}

\smallskip

\begin{center}
{\bf\large   
 THE SEARCH FOR STABLE, MASSIVE, \\ 
 ELEMENTARY PARTICLES\footnote
 {Work supported by Department of Energy contract  DE--AC03--76SF00515.}}
\end{center}

\smallskip
\smallskip

\begin{center}

Martin L. Perl, Peter C. Kim, Valerie Halyo, Eric R. Lee, Irwin T. Lee\\
{\it
Stanford Linear Accelerator Center, Stanford University\\
Stanford, California 94309, U. S. A.\\}
\medskip

Dinesh Loomba \\
{\it
Department of Physics, University of New Mexico\\
Albuquerque, New Mexico 87131, U. S. A.\\}
\medskip

Klaus S. Lackner\\
{\it
Theoretical Division, Los Alamos\\
Los Alamos, New Mexico 87545, U. S. A.\\}

\end{center}

\vfill

\begin{center}
 {\bf\large   
  Abstract }
\end{center}

In this paper we review the experimental and observational searches for stable, massive, elementary particles other than the electron and proton. The particles may be neutral, may have unit charge or may have fractional charge. They may interact through the strong, electromagnetic, weak or gravitational forces or through some unknown force. The purpose of this review is to provide a guide for future searches - what is known, what is not known, and what appear to be the most fruitful areas for new searches. A variety of experimental and observational methods such as accelerator experiments, cosmic ray studies, searches for exotic particles in bulk matter and searches using astrophysical observations is included in this review.

\vfill


\begin{center}

 Submitted to {\it International Journal of Modern Physics A}

\end{center}

\vfill\eject


\setcounter{footnote}{0}
\renewcommand{\thefootnote}{\alph{footnote}}

\section{Introduction and General Considerations}
\noindent

\subsection{Scope of the review}
\noindent
In this paper we review the experimental and observational limits on the existence of {\it stable, massive, elementary particles} - particles at least as massive as the electron, 0.5 MeV/c$^{2}$. We use the term {\it elementary} in a broad sense so that the presently known stable, massive, elementary particles are the electron - a basic elementary particle - and the proton - a composite elementary particle. 

The particles may be neutral, may have unit charge or may have fractional charge. They may interact through the strong, electromagnetic, or weak force, or through some unknown force, as long as the force allows direct or indirect detection of the particle. This review does {\it not} include the particle classes of neutrinos, axions or monopoles.

As the reader already knows, except for the electron and proton, no {\it stable}, massive particles that fit our criteria have been found, although many have been proposed and sought. Our purpose is to provide a guide for future searches - what is known, what is not known, and what appear to be the most fruitful areas for new searches. Therefore this review includes a variety of experimental and observational methods: accelerator experiments, cosmic ray studies, searches for exotic particles in bulk matter, and astrophysical searches for signals from the annihilation of particle-antiparticle pairs.

We have chosen the electron mass as the lower mass limit for this review for two reasons. First, in general, search methods differ above and below about 1 MeV/c$^{2}$. Second, our own experimental interests are in more massive particles, in contrast to neutrinos and axions. 

Particle stability is a loose criterion for us. For example, we are interested in particles whose lifetimes are of the order of the lifetime of the universe so that searches for particles produced in the early universe can be meaningful. However, we are also interested in searches at accelerators for new particles with lifetimes sufficiently long to be directly detected, that is, longer than about 10$^{-6}$ s. Finally, we do not classify bound quarks as stable particles, but we would classify as stable particles free quarks that lived long enough for direct detection. Thus in this review, a stable particle is operationally defined as one that can be observed in isolation and has sufficient lifetime to be detected by the search techniques described in this paper.

While nuclei are not elementary particles, {\it stable} nuclei fit our operational search criteria. In this review we have not explicitly discussed searches for unknown, massive, stable nuclei, however some of the search limits can be applied to the question of the existence of unknown, massive, stable nuclei. 

We have kept this review to a reasonable length by imposing two drastic limitations. First, we give very few details of experimental and observational techniques. Second, while we note and sometimes summarize relevant particle and astrophysical theory, we do not give derivations. In writing a review paper the question always occurs as to the extent of the references. Here also we have chosen to be economical of space. References are given for all quoted searches and for some reviews but we do not give references to general concepts in particle physics or astrophysics, concepts such as supersymmetry theory, early universe cosmology, or cosmic ray phenomenology. 

\subsection{Dark matter}
\noindent
Many present searches for massive elementary particles are motivated by the desire to identify the elementary particle or particles that compose what we call dark matter. Dark matter is thought not to be made up of baryons or charged leptons, possibly to be made up in part of neutrinos, but mostly to be made up of one or more types of unknown elementary particles. This review includes the results of searches for the major classes of proposed dark matter massive candidates, such as WIMPS (weakly interacting massive particles), SIMPS (strongly interacting massive particles) and CHAMPS (charged massive particles).

Our search interests are more general. An unknown, stable, massive particle may exist in numbers much too small to explain the amount of dark matter in the universe. Still, the existence of such a particle would be of great importance. Indeed, one purpose of this review is to urge that a massive particle search effort not cease when the upper limit on that particle's density falls below dark matter density requirements.

\subsection{Massive particle production in the early universe}
\noindent
In models for massive particle production in the early universe, models often motivated by the search for a solution to the dark matter puzzle, it is usually assumed the massive particle X and its antiparticle $\bar{\mbox{X}}$, existed in thermodynamic equilibrium in the very early universe. Then, as the universe cools and expands, the X particles `freeze-out' of equilibrium and become {\it thermal relics}, their density depending upon their mass, $M_X$ and their annihilation cross section $\sigma_{ann}$.

Since the X particles are non-relativistic, most of $\sigma_{ann}$ is in the S wave, leading to $\sigma_{ann}$ inversely proportional to $M_X^2$. The `freeze-out' density, $\rho_X$ is inversely proportional to $\sigma_{ann}$, hence proportional to $M_X^2$. Therefore the larger $M_X$, the larger $\rho_X$, but $\rho_X$  cannot exceed the critical mass density of the universe. This leads to an upper limit on the mass of thermal relics so produced. Griest and Kamionkowski\cite{gri} give the upper limit of $3.4\times10^{5}$ GeV/c$^{2}$, for convenience we use

\begin{equation}
M_{thermal-relic}\leq10^6 \mbox{ GeV/c$^2$}.
\end{equation}

There are proposals for getting around this upper limit on $M_{thermal-relic}$. For example Kolb {\it et al.}\cite{kol} have proposed processes by which particles with masses in the range of $10^{12}$ to $10^{16}$ GeV/c$^{2}$ might be produced in sufficient quantity to explain the abundance of dark matter. If one is searching for massive particles with less than dark matter abundance, these non-thermal relic processes become even more attractive. 

Experimenters who look for massive particles with $M_X>10^6$ GeV/c$^2$ produced in the early universe must balance the time, work, and money necessary for such searches, against how much they believe the proposals for getting around Eq. 1.

\subsection {Galaxy formation and halo particles}
\noindent
During the formation of our galaxy there was an early period when the gravitational force of the galaxy mass controlled the velocity distribution of all individual particles. All particles have the same average velocity, $v_{ave}$, calculated by using the virial theorem, namely:

\begin{equation}
v_{ave}\approx 300 \mbox{ km/s or }\beta_{ave}\approx 10^{-3}. 
\end{equation}

\noindent
Here $\beta=v/c$.

As the galaxy cools, the particle's kinetic energy decreases through collisions with other particles. For an electron, proton or nucleus the kinetic energy is small and most electrons, protons and nuclei condense into ordinary matter. But there are two conditions that must be satisfied so that massive X particles also condense into, or are trapped by, ordinary matter:

a. The interaction of the X with ordinary matter must be strong enough so that the X particle loses its initial high velocity through its collisions with ordinary matter. Electromagnetic interactions or perhaps strong interactions are required. An X particle with only the weak interaction may not be able to lose enough energy through collisions. An example is the WIMP model where WIMPS with masses in the range of 50 to 500 GeV/c$^2$ are assumed to still retain $\beta_{ave}\approx 10^{-3}$.

b. If $M_X$ is very large, say $10^{12}$ GeV/c$^2$, then its initial energy is $10^6$ GeV and the particle will very rarely be trapped in ordinary matter unless it directly hits a large object such as the earth. In that case, depending upon the type of interaction, the X may be stopped in the earth's atmosphere or in the earth's crust or deep in the earth. Of course with sufficient energy the X might pass right through the earth. 

If massive particles of some type are not trapped by ordinary matter and remain moving in random orbits in the galaxy, we call them {\it halo particles}.

The mass boundary between when particles are trapped in ordinary matter and when they remain in the halo of the galaxy depends upon the interactions of the particle. The calculation of the mass boundary is not precise because of the different models used for particle interactions as the galaxy cools. Thus for a positive, unit-charge, massive particle, De R\'{u}jula {\it et al.}\cite{der} give the boundary as $2\times10^4$ GeV/c$^2$, that is, above $2\times10^4$ GeV/c$^2$ these massive particles remain in the halo. On the other hand, Dimopoulos {\it et al.}\cite{dim} give a much higher mass boundary, $10^8$ GeV/c$^2$.

\subsection{Massive particle searches using colliders and accelerators}
\noindent
Searches for stable, massive particles at high energy colliders - e$^{+}$ e$^{-}$, e$^{\pm}$ p, and \ p$^{+}$~p$^{-}$ - have two obvious advantages. First, the search mass range is known and, if a production cross section is assumed, the existence of a hypothetical particle may be directly tested. A simple example is the search for a stable, charged lepton, L$^{\pm}$, using
 
\begin{equation}
\mbox{e}^{+} + \mbox{e}^{-} \to \mbox{L}^{+} + \mbox{L}^{-}.
\end{equation}

A second advantage is that the stability requirement is less restrictive compared to other search methods, usually a lifetime greater than 10$^{-6}$ s is sufficient for direct detection of the particle.

The limitations of collider searches are obvious. The mass range is limited by the available energy. The upper limit to the Tevatron mass search range is several hundred GeV/c$^2$ and that for LEP2 is about 100 GeV/c$^2$. The significance of a null collider or accelerator search is limited by our knowledge of the production cross section. Thus a search for free quarks cannot use conventional quantum chromodynamics (with its confinement hypothesis) to calculate the expected production cross section.

There have also been some interesting studies using a fixed target accelerator. For example Prinz {\it et al.}\cite{jaros} have searched for particles with very small charges.

\subsection{Searches in cosmic rays and halo particles}
\noindent
There are three classes of particles impinging on the earth. The two observationally dominant classes are the primary cosmic rays coming from outside the earth, mostly proton and nucleons, and the secondary particles produced by interactions in the atmosphere such as muons and pions. The primary cosmic ray class is of more interest particularly because of the detection of very high energy particles, those with energies greater than 10$^{19}$ eV. Historically the class of secondary particles has been of great importance because this class was the source of the discoveries of many particles. But, with the steadily increasing energy and intensity of colliders the secondary class is of decreasing interest.

The third class of particles impinging on the earth consists of the assumed halo particles of the galaxy. Interest in this class has, of course, been greatly stimulated by the search for the constituents of dark matter.
 
\subsection{Searches for stable, massive particles in ordinary matter}
\noindent
Stable, massive particles might be introduced into ordinary matter through two different processes. In one process occurring in the course of the cooling of the galaxy, massive particles might condense with baryonic matter to form stars, planets and smaller bodies such as asteroids. As discussed in Sec. 1.4, this process will occur if the particles are not too massive or if in their interaction with ordinary matter they have exceptionally large $dE/dx$.

The other process that could introduce massive particles into ordinary matter would take place once stars and planets are formed. Some types of massive cosmic ray particles or halo particles impinging on the these bodies could lose sufficient energy to become trapped in the ordinary matter of these bodies. An example is the hypothesis that a massive, positive charge particle, X$^{+}$ falling through water might come to rest and form the exotic heavy water molecule HXO.

\section{Searches for Integer Charge, Stable, Massive Particles}
\noindent

\subsection{General considerations}
\noindent
In this section we consider bounds on the existence of integer charge massive particles, X. We assume that the particles can be positively or negatively charged and, although in principle the magnitude could be any integer, we restrict ourselves here to the case of unit charge.

Limits on the existence of these particles come from a variety of experiments. Accelerator searches (Sec. 2.2) have a limited mass reach but are able to detect particles with very small lifetimes. Experiments could also detect present day relic abundance of the X particles from production in some early epoch of the universe. Their subsequent history, which would tell us how best to search for them, depends on their mass and on theories of galaxy formation, as discussed in Sec. 1.4 (see also Refs. \citen{der,dim}). The first of two extreme possibilities is that they remain in the galactic halo maintaining dynamical properties similar to those associated with cold dark matter. The second possibility is that they condense out into the galactic disk and subsequently become trapped within ordinary matter. Searches in cosmic radiation probe the former scenario (Sec. 2.3) and searches in bulk matter (Sec. 2.4) test the latter. In addition, cosmic ray searches can also look for X particles produced and accelerated by high energy astrophysical processes. Finally, indirect limits on X particle properties can be inferred from a number of astronomical observations, as discussed in Sec. 2.5.

Although there has always been interest in searches for trace amounts of particles with characteristics like that of X, the possibility that charged particles could have an abundance large enough to make them viable candidates for dark matter would at first thoughts appear ludicrous. However, as first shown by De R\'{u}jula {\it et al.}\cite{der} and Dimopoulos {\it et al.}\cite{dim}, if the particles are sufficiently massive they can indeed play the role of cold dark matter. This realization led to a revival of interest in searches for charged massive stable particles and much of the work reported below was a consequence of Refs.~\citen{der,dim}. To set the stage for later discussion (Secs.\ 2.3-2.5), \hfill \newline we list here a few properties of dark matter that are generally accepted (see Primack {\it et al.}\cite{pri} for additional details). The first is the estimate of the dark matter density in the local solar neighborhood, $\rho_{DM,local}\sim 0.3$ GeV/cm$^3$. The second is the velocity distribution of dark matter, which is approximately Maxwellian with a cutoff at about 640 km/s (the local escape velocity) and with an rms velocity of about 260 km/s. Thus, the flux of dark matter at earth can be approximated by:

\begin{equation}
\Phi \approx (0.3 \mbox{ GeV/cm}^3)(10^{-3}c)/M_{DM} \approx 10^7 \mbox{cm}^{-2}\mbox{s}^{-1}/(M_{DM}/\mbox{GeV/c}^2).
\end{equation}

\noindent
Here $M_{DM}$ is the mass of a dark matter particle. Note that this flux is larger than the cosmic ray proton flux which is about\cite{gai} 1 cm$^{-2}$s$^{-1}$ for $M_{DM}$=10$^6$ GeV/c$^2$. Finally, an upper limit on the density of dark matter is believed to be about 30\% of the critical density of the universe\cite{cas}, and about ten times more than that of baryons. On scales of our Galaxy, the dark matter resides in a much larger volume (the halo) than the baryons (which are constrained to the disk), hence the low value of 0.3 GeV/cm$^3$ given above for the local neighborhood.

Finally, as will be shown below, most experimental limits indicate that charged massive particles do not fill the budget demands for the density of dark matter. This has led to diminished interest in further searches for charged stable massive particles. However, as recent results are making clear, the energy budget of the universe is not dominated by one species, but is an amalgamation of baryons, neutrinos, exotic dark matter, exotic dark energy, etc\cite{bah}. Could it not also be that the dark matter itself is composed of a spectrum of different particles? We leave it to future experimenters to devote their efforts to searching for X particles whose abundance may be very rare in comparison to known species.

\subsection{Accelerator searches for integer charge, stable, massive particles }
\noindent
Equation 3, the pair production of integer charge particles in electron-positron annihilation, provides the most straightforward way to search for such stable massive particles. The beam energy and the measured momentum provide a determination of the mass of the sought particle, the pair production cross section is known and the LEP2 electron positron collider had a mass search range up to about 100 GeV/c$^2$. Figure 1 shows the results of a search by Acciarri {\it et al.}\cite{l31}, the L3 experiment at LEP. Up to the search limit of 93.5 GeV/c$^2$, no massive particles were found. Similar null searches were carried out by Borate {\it et al.}\cite{al1}, the ALEPH experiment at LEP.

\begin{figure}[htb]
\begin{center}
\epsfig{file=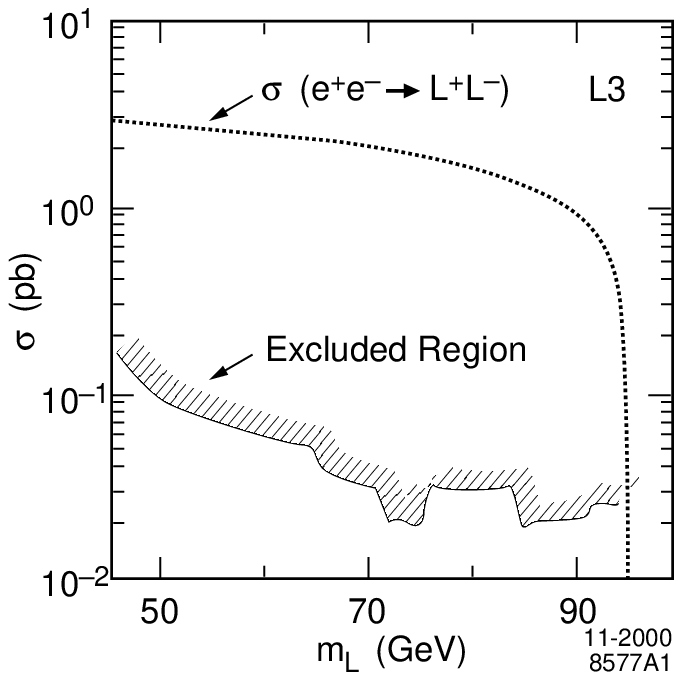,width=10cm}
\caption{Upper limit of the cross section for pair production of stable charged heavy leptons at $\sqrt{s}=133$ to 189 GeV from Acciarri {\it et al.}\protect\cite{l31}. The hatched area indicates the excluded region. The dashed line represents the calculated pair production cross section for heavy leptons at $\sqrt{s}=$ 189 GeV.}
\end{center}
\end{figure}

\vfill
\newpage

P. Abreu {\it et al.}\cite{de1}, the DELPHI experiment at LEP, carried out a
more general search at 189 GeV for integer charge, stable, massive particles produced in events with two or three charged particles. The motivation was to look for the supersymmetric partners of the muon and tau, assuming these partners to be stable. No signal was found in the search mass range of 2 to 80 GeV/c$^2$, but in this case the significance of the search depends upon the assumed cross section.

Searches for integer charge, stable, massive particles at the Tevatron reach to larger masses than searches at LEP2. Abe {\it et al.}\cite{abe} looked for pair production of particles with masses up to the order of 500 GeV/c$^2$. No evidence for such particles was found\cite{abe}. However, proton-antiproton production cross section calculations do not have the certainty of electron-positron electromagnetic pair production cross sections, hence the significance of the null results depends upon the model used for the production cross section, Fig. 2.

\begin{figure}[htb]
\begin{center}
\epsfig{file=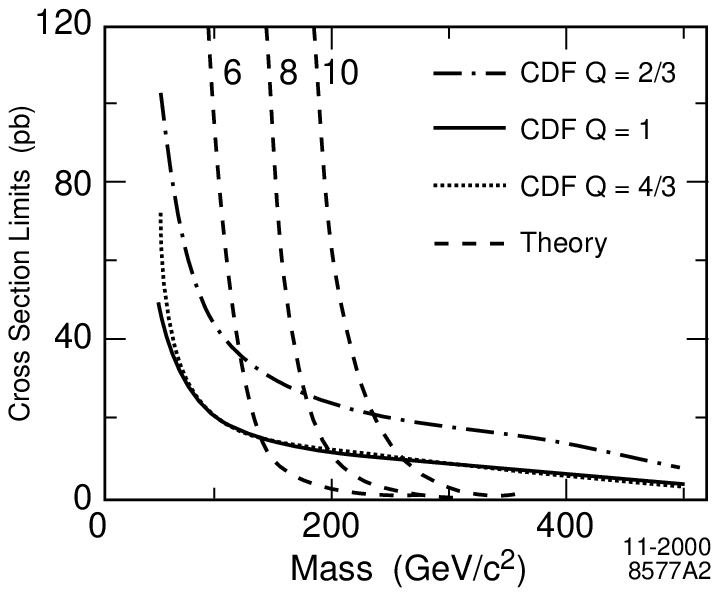,width=10cm}
\caption{The cross section upper limits (95\% C.L.) for the pair production of stable charged fermionic colored particles for $Q=1$, $Q=\frac{2}{3}$ and $Q=\frac{4}{3}$ from Abe {\it et al.}\protect\cite{abe}using the Tevatron. Also shown are the theoretical cross sections for the production of fermionic color sextets(6), octets(8) and decuplets(10).}
\end{center}
\end{figure}

\subsection{Cosmic rays, halo particles, and searches for integer charge, stable, massive particles}
\noindent
Before discussing experimental limits on X particles in cosmic rays and the galactic halo, we turn to some preliminary considerations. First, as noted in Sec.~2.1, much of the interest in X particles has been generated in the hope of identifying the X as exotic dark matter. This simplifies the search because, as discussed in Sec.~1.4, the conditions for the X particles to reside in the halo (a dark matter requirement) are that they have a large mass, M$_X\sim 10^5$ GeV/c$^2$, and that their velocity be close to\cite{der,dim} $\beta\sim 10^{-3}$. Therefore one looks for slow, charged particles whose velocities remain constant as they pass through the detector. In contrast, a general search for a rare X in the cosmic radiation must measure a combination of particle parameters, such as mass, velocity, and charge, in order to rule out the large background of naturally occurring isotopes. Ultimately, however, it may be impossible to distinguish the origin of a detected X particle: it may come from a relic population circulating in the halo, a primary cosmic ray made in an astrophysical source or a secondary cosmic ray created in a high-energy collision in the earth's atmosphere. Thus, in our discussion below we will not distinguish between these but will collectively refer to them as cosmic ray particles.

A second consideration pertains to the identification of X as it passes through the detector. Over their history the X$^+$ and X$^-$ particles may take different guises. For example, the X$^+$ could acquire an electron to form an isotope of hydrogen X$^+$e$^-$. Similarly, the X$^-$ could have bound to either a proton to form a neutral composite, X$^-$p$^+$, or to an alpha particle which later captures an electron to form a heavy hydrogen-like atom, X$^-\alpha^{++}$e$^-$, similar to the X$^+$e$^-$. The X$^+$e$^-$ and X$^-\alpha^{++}$e$^-$ are collectively denoted as CHAMPS by De R\'{u}jula {\it et al.}\cite{der} whereas the X$^-$p$^+$ are termed neutraCHAMPS. On entry into the atmosphere, the neutraCHAMPS very quickly exchange their proton for $^{14}$N. Thus only satellite or balloon searches have to worry about missing a neutraCHAMP because of lack of a signal and, conversely, probably only these experiments can distinguish neutraCHAMPS from charged CHAMPS by observing an exchange occurring in their detector (e.g., Barwick {\it et al.}\cite{bar}, discussed below in  Sec. 2.3.1).

Finally, we consider the upper bound on the mass of X. As discussed in Sec. 1.3, the upper bound on the mass of any particle that is produced in the early universe, regardless of its interactions, depends on our understanding of the physics of this epoch. Although the accepted view holds that M$_X\leq 10^6$ GeV/c$^2$, a number of recent works, for example Chung {\it et al.}\cite{chu}, have described scenarios which allow this limit to be extended, even to the GUT scale. Chung {\it et al.}\cite{chu} resurrect the possibility that not only could dark matter particles be supermassive but that they could also be charged. We find, in fact, that the latter possibility can be ruled out by considering the results of the MACRO experiment. We discuss this below~in~Sec.~2.3.2.

\subsubsection{Balloon and satellite search limits}
\noindent
Although no satellite or balloon experiment was designed to explicitly search for CHAMPS or neutraCHAMPS, limits on their abundances have been obtained using data from three such experiments that were originally intended for cosmic ray studies. In all three cases the data were obtained and reanalyzed with the intent to test the charged dark matter hypothesis proposed by Refs. \citen{der,dim}. From Eq. 4 we note that for low mass values, $M_X< 10^6$ GeV/c$^2$, the dark matter flux at the top of the atmosphere exceeds that of cosmic rays. This fact is utilized in each of these studies to rule out X masses for which the expected dark matter flux is higher than observed. Note that the range of excluded X masses also has a lower limit. This is because at small enough X masses the kinetic energy of the particle is no longer sufficient for it to penetrate both the solar wind and the minimum number of detector elements required by the analysis selection criteria.

The first limit on the X abundances in cosmic rays\cite{dim} was set by using data from track-etch detectors employed in the University of Chicago experiment on the Pioneer spacecraft. These detectors are ideal for searching for X particles in cosmic rays because their restricted sensitivity to particles with charge to velocity ratios of $Z/\beta\geq 14$, for $\beta>10^{-2}$ naturally eliminates the dominant background of fast protons and alpha particles. The response of these detectors to CHAMPS with $\beta\approx 10^{-3}$ is estimated\cite{dim} to be comparable to that of cosmic rays with $Z/\beta\approx 16$. Ref. \citen{dim} also notes however that calibrations of track-etch detectors are limited to charged particles with $\beta \geq 10^{-2}$. Nevertheless, with a number of caveats, dark matter being entirely composed of CHAMPS of mass $M_X \leq 10^7$ GeV/c$^2$ is ruled out. However, no limits were obtained on the neutraCHAMP abundance.

The second analysis, Barwick {\it et al.}\cite{bar} employed track-etch data from a balloon flight and set bounds on both CHAMP and neutraCHAMP masses. This more careful analysis includes detailed calculations of CHAMP energy loss, which appear to confirm the sensitivity of track-etch detectors to low velocity massive charged particles. Barwick {\it et al.}\cite{bar} rule out dark matter CHAMPS in the mass range $3.5 \times 10^2 < M_X < 8.6 \times 10^7$ GeV/c$^2$. In addition they also rule out neutraCHAMPS as dark matter for $10^2 < M_X < 4 \times 10^3$ GeV/c$^2$.

The last analysis, by Snowden-Ifft {\it et al.}\cite{sno} uses data from the IMP 8 satellite to rule out CHAMP dark matter in the mass range $2.4 \times 10^3 < M_X < 5.6 \times 10^7$ GeV/c$^2$. Limits from all three analyses are shown in Fig. 3.

\vfill\newpage

\begin{figure}[htb]
\begin{center}
\epsfig{file=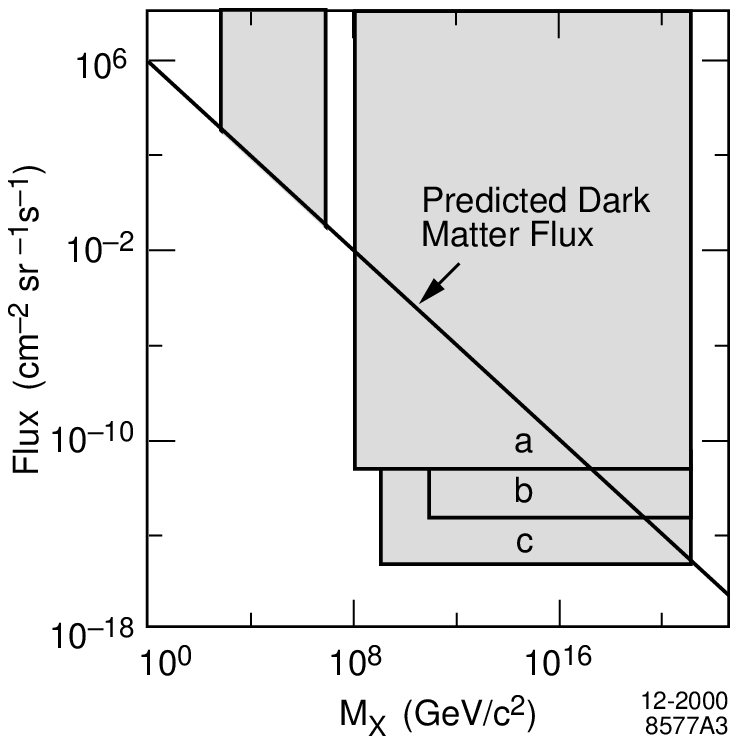,width=10cm}
\caption{Observed upper limits on the flux of CHAMPS (charged massive particles) versus the CHAMP mass, $M_X$. The diagonal line gives the predicted dark matter flux as a function of $M_X$. The excluded, shaded region on the left comes from the observations discussed in Sec. 2.3.1. The excluded, shaded regions marked a, b and c come from (a) Eq. 7, (b) Eq. 5, and (c) Eq. 6 in Sec. 2.3.2.}
\end{center}
\end{figure}

\subsubsection{The MACRO experiment searches for integer charge, stable, massive particles}
\noindent
Of the ground- and underground-based experiments, two place very strong limits on the flux of X particles in cosmic rays: MACRO, based deep underground and a precursor to MACRO, placed on the surface of the earth.

The MACRO detector\cite{ahl}, based deep underground in Gran Sasso, Italy, was primarily designed to search for magnetic monopoles. In order for the X particle to be detected by MACRO, which has a minimum overburden of 3300 m water equivalent, it must be highly penetrating and, therefore, must have a large kinetic energy. This places a lower limit on the mass of X. For example, if X has a typical halo particle velocity of $\beta \sim 10^{-3}$ at the surface of the earth, it must have a minimum mass of $M_{X,min} \approx 10^{11}$ GeV/c$^2$ to reach the MACRO detector. Here we have used a mean energy loss\cite{der,mue} of $dE/dx \approx$ 150 MeV/g/cm$^2$ valid over $\beta \sim 10^{-4}$ to $10^{-3}$. The MACRO detector consists of a series of subdetectors that are used both individually and cumulatively to set flux limits for monopoles and a number of other rare exotic particles\cite{amb}. Although limits on integer charge particles, such as CHAMPS, are not quoted explicitly in Ambrosio {\it et al.}\cite{amb} they should not be too different from their limits on the monopole flux obtained using the liquid scintillator subdetector system. Both CHAMPS and monopoles are slow, highly ionizing, and penetrating and should give similar responses from scintillators. In fact, MACRO was optimized in part by utilizing the measured light yield of slow, $\beta \approx 2.5 \times 10^{-4}$ protons in scintillators\cite{fic}, which should be comparable to that of CHAMPS and is estimated to be larger than that due to bare GUT monopoles for $10^{-4} \leq \beta \leq 2 \times 10^{-3}$. The triggers that MACRO employs for monopoles in the velocity range $10^{-4} \leq \beta \leq 10^{-1}$ should also have comparable efficiency for triggering CHAMPS. Therefore, we use the Ambrosio {\it et al.}\cite{amb} scintillator derived 90\% C.L. upper limit on monopole flux in the velocity range $\beta\sim 10^{-4}$ to $10^{-3}$ as a conservative bound on the flux of X particles with mass M$_X \geq 10^{11}$ GeV/c$^2$ in the same velocity range:

\begin{equation}
\Phi \leq 4 \times 10^{-14} \mbox{cm}^{-2}\mbox{sr}^{-1}\mbox{s}^{-1}.
\end{equation}

\noindent
For faster particles, $\beta \sim 1.2 \times 10^{-3}$ to $10^{-1}$, a better limit is given: 

\begin{equation}
\Phi \leq 2.6 \times 10^{-16} \mbox{cm}^{-2}\mbox{sr}^{-1}\mbox{s}^{-1}
\end{equation}

\noindent
which still overlaps with the high velocity tail of the expected dark matter flux. In this case, however, the mass reach is lowered to as much as M$_X \geq 10^9$ GeV/c$^2$ because of the higher kinetic energies of the particles. Both of these limits are shown in Fig. 3 together with the flux expected from dark matter.

As mentioned earlier, this important limit on the flux of X particles pertains directly to the recent proposal by Chung {\it et al.}\cite{chu} for supermassive dark matter candidates. They alluded to the possibility that such particles could be charged. However, if $M_X \geq 10^{11}$ GeV/c$^2$, the above limits surely rule out this possibility.

Results from a precursor experiment to MACRO placed at the surface of earth\cite{bari}, also with the intent to search for magnetic monopoles, can be used to place limits on X particles in the mass range M$_X \geq 10^8$ GeV/c$^2$. Below this mass a particle with velocity $\beta \approx 10^{-3}$ at the top of the atmosphere ranges out before hitting the ground. Their 90\% C.L. upper flux limit for particles with velocity $5 \times 10^{-4} \leq \beta\leq 2.7 \times 10^{-3}$ is:
 
\begin{equation}
\Phi \leq 4.7 \times 10^{-12} \mbox{cm}^{-2}\mbox{sr}^{-1}\mbox{s}^{-1}.
\end{equation}

As shown in Fig. 3 this limit, together with those from MACRO and the satellite and balloon limits (Sec. 2.3.1), essentially close the fully allowed mass range for charged dark matter of the type considered by Chung {\it et al.}\cite{chu}

\subsection{Searches in ordinary matter for integer charge, stable, massive particles }
\noindent

\subsubsection{Searches for massive particles in water}
\noindent
It is an attractive idea that a stable, massive, positively charged particle, X$^+$, falling onto and through oceans and lakes, will form the heavy water molecule HXO. Table~1 lists the four experiments that have searched for HXO. No evidence was found for an X$^+$ in the mass ranges given in the table. The upper limits on the concentration of HXO in H$_2$O are given in Fig. 4.

\begin{figure}[htb]
\begin{center}
\epsfig{file=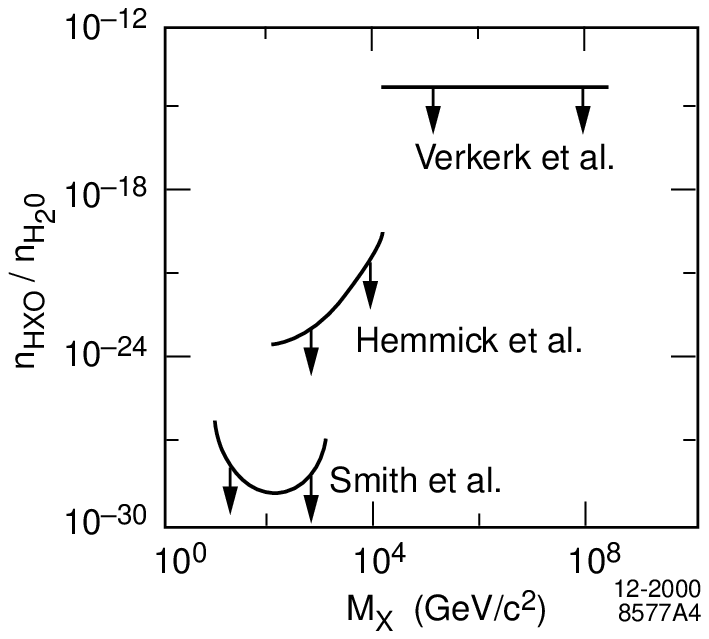,width=9cm}
\caption{Observed upper limits on $n_{HXO}$/$n_{H_{2}O}$, the concentration of HXO in H$_2$O versus the mass of a unit charge particle X. Limits are from Vekerk {\it et al.}\protect\cite{ver}, Hemmick {\it et al.}\protect\cite{hem1} and Smith {\it et al.}\protect\cite{smi1}.}
\end{center}
\end{figure}

\subsubsection{Searches for massive particles in isotopes}
\noindent
A massive particle with negative integer charge, X$^-$, can electromagnetically bind with a very small Bohr orbit to a nucleus\cite{hem1}. If the nucleus has charge Z, this gives an anomalously heavy, usually called superheavy, isotope with charge Z-1. There have been several searches for such isotopes.

Hemmick {\it et al.}\cite{hem1} searched for superheavy isotopes of lithium, beryllium, boron, carbon, oxygen and fluorine in a mass range up to about 10$^5$ GeV/c$^2$. No superheavy isotopes were found with the upper limits on the number of superheavy isotopes per nucleon ranging from $10^{-10}$ to $10^{-17}$. Similar upper limits were found by Turkevich {\it et al.}\cite{tur} for carbon and Dick {\it et al.}\cite{dic} for sodium.

\subsubsection{Comparison of bulk matter limits with cosmic ray limits}
\noindent

\begin{table}[t]
\begin{center}
\vspace*{12pt}
\caption{Searches for stable, massive, positively charged particles in water. The upper limits on the concentration of HXO in H$_2$O are given in Fig. 4.}
\vspace*{10 pt}
\begin{tabular}{|l|l|l|} \hline
\bf Concentration Method & \bf Detection Method & \bf Experiment \\
\hline \hline
electrolysis & mass spectrometer & Smith {\it et al.}\cite{smi1}\\
\hline
see paper & mass spectrometer & Hemmick {\it et al.}\cite{hem1}\\
\hline
ultracentifuge & spectral analysis & Verkerk {\it et al.}\cite{ver}\\
\hline
none & mass spectrometer & Yamagata {\it et al.}\cite{yam}\\
\hline \hline
\end{tabular}
\end{center}
\end{table}

The relative fraction of X particles in the halo versus condensed matter in the disk (e.g., stars, planets, etc.) is believed to be large if the mass of X is much larger than 10$^5$ to 10$^8$ GeV/c$^2$ (Sec. 1.4).  This, however, does not mean that a significant enrichment could not have occurred in the earth from stopping X particles over many years. Therefore, one must weigh the benefits of large throughput experiments in bulk matter, which may contain very small abundances of X particles, over cosmic ray experiments which have small areas and exposure times but probe the primary X particle flux. For this purpose it would be useful to convert the upper bounds on the flux of X particles from Sec. 2.3 into an enrichment of ordinary matter on earth. Here we describe a simple, illustrative model that does this.

The assumptions used in the model are extremely simple. The enrichment of earth by X particles is assumed to occur continuously over its age, 4 Gyr, during which time geophysical or geochemical processes are ignored. The flux of X particles and their corresponding masses are taken from the best limits set by experiments described in Sec.~2.3 (see Fig.~3). The range of these particles in the earth is calculated assuming a constant velocity of $\beta\approx 10^{-3}$ upon entry, and a mean energy loss of $dE/dx=150$ MeV/g/cm$^2$. The particles are assumed to stop uniformly down to a depth given by their range. Finally, the fraction of X particles in ordinary matter, $n_X/n_N$, is calculated at the surface and the result is compared in Fig. 5 to the results of bulk matter searches given in Fig. 4.

\begin{figure}[htb]
\begin{center}
\epsfig{file=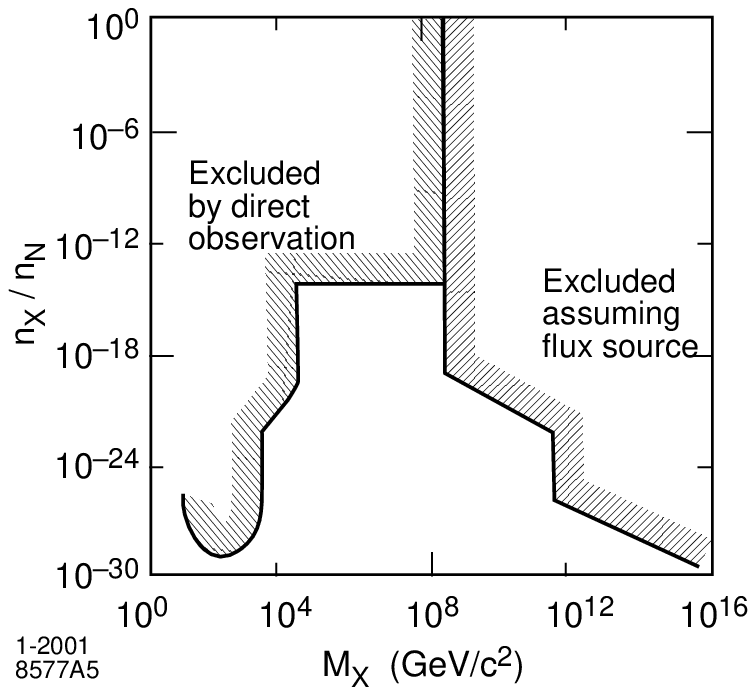,width=10cm}
\caption{Upper limits on the fraction of X particles in ordinary matter, $n_X/n_N$ where X is a unit charge particle. The left excluded region is from Fig. 4. The right excluded region is calculated from the upper limits on the flux of X particles at the earth's surface using the model in Sec. 2.4.3. This right exclusion region assumes that all X particles in the earth come from the flux of X particles at the earth's surface.}
\end{center}
\end{figure}

\subsection{Limits from astrophysics on the existence of integer charge, stable, massive particles }
\noindent
A number of constraints can be placed on the nature of integer charge massive stable particles from astrophysics and cosmology. One cosmological constraint, the upper limit on the mass of a particle produced in the early universe, M$_X\leq 10^6$ GeV/c$^2$, has already been discussed in Sec. 1.3. Another constraint, discussed in Sec. 2.1, is that the X particle density cannot exceed the upper limit on the density of exotic dark matter which is approximately 30\% of the energy density of the universe. The local density of this component is believed to be of order 0.3 GeV/c$^2$ which sets an upper limit on the flux on X particles at earth (see Eq. 4). Lower bounds on $M_X$ have also been set based on theories of galactogenesis and assuming that X comprises 100\% of dark matter as given in Ref. \citen{der,dim}. However, as discussed in Sec. 1.4, their limits, $2\times10^4$ GeV/c$^2$ and 10$^8$ GeV/c$^2$, disagree by a large margin. If we adopt the conservatively calculated higher limit of Dimopoulos {\it et al.}\cite{dim}, and believe the upper bound $M_X\leq 10^6$ GeV/c$^2$ set from cosmology, this essentially rules out the CHAMP dark matter scenario. Below we outline two more examples of astrophysical bounds to give a flavor of this type of inference.

Chivukula {\it et al.}\cite{chi} rule out CHAMP dark matter based on measurements of cooling rates of clouds of hydrogen gas in the interstellar medium. Their argument relies on noting that in equilibrium the heating of the interstellar gas from interactions with CHAMPS cannot exceed the cooling rates. This places an upper bound on the strength of interactions between halo dark matter and the interstellar medium. By estimating the elastic CHAMP-hydrogen interaction cross section and using the local dark matter density, they find that $M_X >10^6$ GeV/c$^2$.  This limit only applies to CHAMPS (i.e., X$^+$e$^-$ and X$^-\alpha^{++}$e$^-$) and not to neutraCHAMPS.

Gould {\it et al.}\cite{gou} also rule out CHAMP dark matter but in a large mass range, $10^2\leq M_X \leq 10^{16}$ GeV/c$^2$. They argue that the hypothetical dark matter CHAMPS (X$^+$e$^-$ and X$^-\alpha^{++}$e$^-$) would accumulate in protostellar clouds with a sufficient concentration to disturb latter phases of stellar evolution. Specifically, in the case where a neutron star is the endpoint of a star's life, the CHAMPS drift to its center and form a black hole which quickly destroys its host via accretion. For dark matter CHAMPS in the mass range given above, Gould {\it et al.} calculate that the destruction of the neutron star would occur on time scales less than 10 yr, which is much shorter than the ages of old neutron stars ($\geq$1 Gyr).

\section{Searches for Neutral, Stable, Massive Particles}
\noindent
Weakly interacting massive particles (WIMPS) are considered one of the leading candidates for cold dark matter\cite{pri,jun,ros}. Of the various types of WIMPS, the lightest supersymmetric particle (LSP, e.g., neutralino) with mass between 1 GeV/c$^2$ and 1 TeV/c$^2$ and interaction cross section with ordinary matter $\sigma < \sigma_{weak}$, is amongst the favored\cite{ros}. In Sec. 3.1 we briefly review the experimental status of dark matter WIMP searches and show the current best limits. In Sec. 3.2 we discuss another candidate for dark matter, strongly interacting massive particles (SIMPS), which, except for a small region of mass versus cross section space, have~been~ruled~out.

\subsection{Searches for WIMPS - weakly interacting massive particles}
\noindent
There have been numerous reviews on the subject of WIMPS\cite{pri,jun,ros,moral,berg,bau,ram}and descriptions of various experiments and their status\cite{moral,pro}. Here we discuss some general features of dark matter WIMP searches and briefly review the most sensitive current experimental limits. Ellis {\it et al.}\cite{ell} have reviewed accelerator searches and limits. There are two basic methods for detecting dark matter WIMPS: the direct and the indirect search methods. In the direct method (Sec. 3.1.1) the signal of a WIMP interacting in the detector (e.g., from WIMP-nucleon elastic collision) is directly measured. In the indirect method (Sec. 3.1.2) it is the products of WIMP-antiWIMP annihilation (e.g., gamma rays, positrons, anti-protons, or neutrinos) which are detected. Presumed sources can be in the galactic halo or in specific sites where WIMPS can gravitationally accumulate over time, such as at the center of the Galaxy, earth or sun.

\subsubsection{Direct searches for dark matter WIMPS}
\noindent
In this category there are currently over twenty experiments that are in various stages of planning, construction, or running\cite{ros,abu}. One experiment, DAMA, has in fact claimed a positive signal, which they find to be consistent with a WIMP mass of $52^{+10}_{-8}$ GeV/c$^2$ at 4$\sigma$ C.L. However, a second experiment, the Cryogenic Dark Matter Search (CDMS), finds a null result even though their sensitivity is close to that of DAMA's. Below we review the results of these experiments but, first, we discuss some general features of direct dark matter WIMP searches.

The differential interaction rate of WIMPS in a detector, in units of rate per unit energy per unit detector mass, is given by\cite{jun}:

\begin{equation}
\frac{dR}{dQ}=\frac{\rho_{DM} \sigma}{2m_W \mu^2}F^2(Q)\int_{v_{min}}^\infty \frac{f(v)}{v} dv.
\end{equation}

\noindent
Here $\rho_{DM}=0.3$ GeV/cm$^3$ is the local density of dark matter (the assumed WIMP density), $m_W$ is the WIMP mass, $\sigma$ is the WIMP-nucleon elastic interaction cross section, $\mu$ is the WIMP-nucleon reduced mass, $F(Q)$ is a nuclear form factor, $v_{min}$ is the minimum detectable WIMP velocity, and $f(v)$ is the distribution of WIMP velocities relative to the detector. Also

\begin{equation}
v_{min}=\left(\frac{E_{thr}m_N}{2\mu^2}\right)^{\frac{1}{2}}
\end{equation}

\noindent
where $E_{thr}$ is the detector energy threshold and $m_N$ is the target nucleus mass. Note that the form factor $F(Q)$ depends on WIMP-nucleus coupling, which can be spin-dependent or spin-independent. In general the spin-independent coupling dominates and it is from results of experiments probing this interaction that the most sensitive bounds exist. See Fig. 5 of Ref. \citen{bau}   for limits from recent spin-dependent results. 

The unknowns in Eq. 8 which are sought are $m_W$ and $\sigma$. The astrophysical quantities $\rho_{DM}$ and $f(v)$ are determined from models of the galactic halo in which the WIMPS are assumed to be distributed in a non-rotating isothermal sphere with a velocity distribution close to Maxwellian\cite{pri,jun} (see Sec. 2.1). However, see Evans {\it et al.}\cite{eva}, Green\cite{green} and Ullio and Kamionkowski\cite{ull} for discussions of deviations from the standard halo model. In this scenario the sun moves through the WIMP ``wind" with an average velocity\cite{jun} of 220 km/s, whilst the earth's velocity relative to the wind varies by a few percent as it revolves around the sun. This variation should result in a seasonal (sinusoidal) modulation of the event rate together with a variation of the energy spectrum. The event rate should be a maximum when the earth's motion is aligned with the sun's (June) and a minimum when it is anti-aligned (December). The effect, however, is only of order several percent and to be seen it requires a large detector to acquire the necessary statistics. Nevertheless, if seen, it is one of several important criteria that can be used to separate signal from background\cite{pri}. These seasonal effects on the rate and spectrum are, in fact, what DAMA observes and are the main basis for their claim.

To understand detector requirements it is important to know the rates expected from dark matter WIMPS. Typical rates predicted from supersymmetry models range from $10^{-5}$ to 10 events/kg/day, for a deposition of energy between 1 and 100 keV in the detector\cite{ros}. If one includes cosmological constraints, such as requiring WIMPS to fill the dark matter budget, then one gets rates $\leq 10^{-2}$ events/kg/day, which are about an order of magnitude below current detector sensitivities. Given the rarity of WIMP interactions it is important to understand and use specific signatures that can be used to eliminate backgrounds and bolster claims\cite{pri,jun,ros}. Two of these, the modulations of event rates and corresponding variations of the energy spectrum, have already been mentioned. Another important WIMP signature can be obtained if the directional information of the nuclear recoil in the detector is available. The nuclear recoil is anisotropic due to the seasonal modulation of the flux discussed  above. However, the recoil signature is larger and requires fewer statistics to resolve\cite{copi}. The DRIFT experiment is the only one currently running which employs this signature\cite{leh}. Also important are signatures which specifically distinguish the WIMP signal from cosmic ray backgrounds and radioactivity backgrounds which are by far the largest and most insidious. The reduction of this background requires using a combination of radio-pure detector elements, shielding, and the ability to distinguish WIMP interactions in the detector from those of the background. The WIMPS (and neutrons) deposit a much larger fraction of their energy as heat (via nuclear recoils) whereas interactions from backgrounds such as gammas, electrons, and alpha particles result in ionization from electron recoils. The CDMS experiment, for example, measures both heat and ionization deposition in their detectors, using the ratio of the two to segregate the background events from the neutron or possible WIMP events. Other techniques for background reduction are given by\cite{moral,abu,ull,gol}. Other signatures of WIMP detection can be found in Ref. \citen{pri}.

The current experimental state is dominated by the recent discovery claim from the DAMA experiment\cite{bern} at the Gran Sasso National Laboratory of I.N.F.N. DAMA employs large radiopure NaI crystal scintillators as their detectors. They use 9 NaI crystals, 9.7 kg each, allowing them to obtain 57,986 kg-day worth of statistics spanning four different yearly cycles (which are also reported individually as DAMA/NaI-1 through DAMA/NaI-4). From calibration data they determined that nuclear recoils from neutrons produced faster signals in the NaI detectors than electron recoil pulses from gammas or electrons. This timing difference is used in a pulse shape analysis for background discrimination after which their background rate is lowered to 0.5-1 events/keV/kg/day in the 2-3 keV part of the spectrum, and about 2 events/keV/kg/day in the 3-20 keV region\cite{moral}. Their claim of finding a WIMP ($m_W = 52^{+10}_{-8}$ GeV/c$^2$ and $\xi\sigma =7.2^{+0.4}_{-0.9}\times 10^{-6}$ pb, at 4$\sigma$ C.L.) is based on their observing a modulation in their background-subtracted rates consistent with that due to the seasonal modulation signature discussed above. Figure 2 of Ref. \citen{bern} shows the modulation signature. Here $\xi = \frac{\rho_W}{\rho_{DM}}$is the fraction of the dark matter density made up of WIMPS; it appears often in the dark matter literature. Thus the cross section is uncertain by the factor $\xi$. 

This result, however, has met with some skepticism and criticism, most of which is centered on the shape of the background-subtracted spectrum\cite{ros,gerb}. In addition, raising further questions\cite{ros,mor}, other NaI experiments employing similar pulse shape discrimination have encountered a class of unexplained events with pulse times close to, or even shorter than, those expected from neutrons or WIMPS. Finally, the CDMS experiment, with sensitivity similar to that of DAMA, has recently released results that are inconsistent with DAMA's allowed $m_W-\xi\sigma$ region\cite{abu,gol}.

The CDMS experiment\cite{abu,gol} is currently housed in the Stanford Underground Facility. CDMS has the ability to measure both phonons and ionization on an event-by-event basis. Using the ratio of energy-loss in phonons to ionization, which for nuclear recoils is lower by a factor of 3 than for electron recoils, the CDMS experimenters are able to achieve a background reduction of greater than 99\%. This, together with the use of radiopure detector elements and care in shielding, has allowed CDMS to reach background rates below 0.1 events/keV/kg/day in the 10-20 keV region of the spectrum\cite{moral2}. Thus, although the small size of the CDMS apparatus limits exposure to only 10.6 kg-days, the ability to efficiently reject backgrounds makes the CDMS experiment competitive. From the data set achieved in this exposure, they report 13 events resulting from nuclear recoils between 10 and 100 keV\cite{abu}. Although this rate is similar to that expected from DAMA's claimed signal region, the CDMS analysis of these events finds the events to be consistent with neutrons\cite{abu}. Thus the findings of the CDMS experimenters rule out DAMA's signal region at greater than 75\% C.L. In addition, the CDMS experimenters find their data to be incompatible with the DAMA modulation signal shown in Fig. 2 of Ref. \citen{bern} at greater than 99.98\%. The CDMS and DAMA results are shown in Fig. 6.

The peculiar situation existing today in the field of dark matter searches surely will not last. CDMS will soon move to the Soudan Mine where the background rate should go down to $10^{-4}$ to $10^{-3}$ events/keV/kg/day\cite{moral2}. The resulting increase in the expected sensitivity of CDMS at Soudan, shown in Fig. 6, will allow it to probe the $m_W-\xi\sigma$ space some 2 orders of magnitude below DAMA's allowed region. Other experiments coming on line will have similar or even better reach and together they will soon be in a position to verify or rule out the DAMA result.

\begin{figure}[htb]
\begin{center}
\epsfig{file=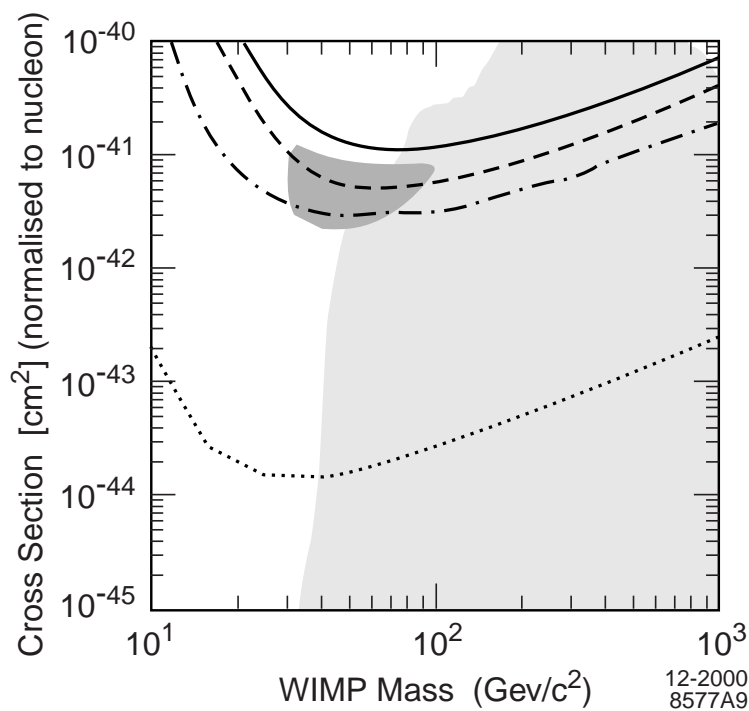,width=10cm}
\caption{Current observations and limits on the WIMP-nucleon spin-independent cross sections as a function of mass. See the text for explanation of the curves and areas.}
\end{center}
\end{figure}

\vfill
\newpage

In Fig. 6, the top three curves are upper limits from: (solid) the Heidelberg-Moscow experiment\cite{baud}, (dash-dash) the 1996 DAMA experiment\cite{bern} and (dot-dash) the CDMS experiment\cite{abu,gol}. The limit from the Heidelberg-Moscow experiment\cite{baud} is shown because it has achieved the lowest background rates directly from raw data. The dark shaded area is the observed 3$\sigma$ signal from the DAMA experiment. The light shaded area shows the theoretical parameter space for various supersymmetric models for WIMPS. The dot-dot curve shows the {\it projected} upper limit that may be reached by the CDMS experiment when it is in the Soudan mine. The figure was obtained from the limit plots web tool provided by R. Gaiskell and V. Mandic at http://cdms.berkeley.edu/limitplots. At this web site, combinations of current and projected future experimental limits for both spin-dependent and spin-independent cross sections can be viewed.

\subsubsection{Indirect searches for dark matter WIMPS}
\noindent
The detection of dark matter through its annihilation signal from various astrophysical sites is the concept behind a number of experimental programs\cite{berg,moral2}. WIMPS should accumulate gravitationally in the center of the sun, and the earth, increasing the annihilation rate relative to that from WIMPS in the halo. WIMPS could also accumulate in the center of the galaxy\cite{gon}. In all three cases the resulting enhancement in the directional anisotropy from the decay products can be utilized as a WIMP signature. WIMP annihilation in the halo, although carrying much less directional information, may produce excesses of gamma rays, positrons or anti-protons above that expected from standard astrophysical processes\cite{berg2,pros,ull2,coutu}. There have been a number of reports of features in positron\cite{coutu} or anti-proton spectra in cosmic rays\cite{buff} which have been interpreted as coming from WIMP annihilation in the halo, but none has been confirmed\cite{berg}. 

Gamma rays and neutrinos provide a somewhat clearer signal because, unlike cosmic rays, they carry directional information. Furthermore, for gamma rays the spectrum can be probed for features from monoenergetic photons produced in WIMP annihilation in various scenarios. For example, if $WW\rightarrow\gamma\gamma$ then $E_{\gamma}\approx m_W$; see Ref. \citen{berg2} for further examples. In fact, simulations shown in Ref. \citen{berg} of the sensitivities of the next generation ground and space-based gamma ray detectors to WIMP annihilation products $\gamma\gamma$ and Z$\gamma$ indicate that competitive constraints could be placed in supersymmetric model parameter space.

The possible accumulation and subsequent trapping of WIMPS in the sun and earth are the best hope for using neutrinos as probes of WIMP annihilation\cite{berg}. High energy neutrinos, the final products in a majority of WIMP annihilation schemes, can provide a clear WIMP signal if observed coming from the sun and the earth. Calculations\cite{berg,moral2} of event rates in detectors from these energetic neutrinos indicate that a detector of area 1 km$^2$ could begin probing interesting regions of supersymmetric parameter space. A number of large neutrino detectors (e.g., AMANDA, Baksan, MACRO, and Super-Kamiokande) have reported flux limits from point sources, although none has shown an excess from any of those considered above. Reference \citen{super} reviews recent results. The advent of next generation large neutrino detectors should provide competitive and complementary limits to direct searches for WIMP dark matter.

\subsection{SIMPS}
\noindent
In the mid-1980's it was noted that while attention was centered on WIMPS, strongly interacting massive particles (SIMPS) could also satisfy the dark matter \newline hypothesis\cite{good,gold}. In fact, their interaction cross sections with matter lie anywhere from $\sigma > \sigma_{weak}$ up to a barn. Numerous candidates were suggested\cite{good,gold}, including both elementary particles and bound states, and it was suggested that they could be detected using detectors similar to those used to search for WIMPS\cite{good}. Soon after the CHAMPS hypothesis (see Sec. 2), Starkman {\it et al.}\cite{stark} applying a similar analysis, calculated existing constraints on the cross sections of dark matter SIMPS for masses greater than about 1 GeV/c$^2$. They found a few interesting open mass ranges: $m<10^5$ GeV/c$^2$ with cross sections greater than $10^{-23}$ cm$^2$; $10^5< m<10^8$ GeV/c$^2$ with cross sections greater than $10^{-27}$ cm$^2$; and $m>10^{10}$ GeV/c$^2$ with a very broad range of cross sections that extends to below $10^{-32}$ cm$^2$. In addition, for a number of these regions, the viability of SIMPS as dark matter depends on whether or not there is SIMP-antiSIMP symmetry in nature. See Figs. 1-3 of Starkman {\it et al.}\cite{stark} for details. Below we briefly summarize the experimental efforts that have been undertaken to close the SIMP dark matter windows.

A combination of balloon-borne, ground-based, and underground experiments have essentially closed the $10^5<m<10^8$ GeV/c$^2$ mass range for dark matter SIMPS\cite{mcg,bern2,derb}. In addition, the underground experiment (DAMA), which is ideal for searching for rare penetrating particles, has closed much of the high mass $m>10^{10}$ GeV/c$^2$, $\sigma>10^{-30}$ cm$^2$, region\cite{derb}. Mohapatra and Nusinov\cite{mohap} show that other WIMP searches, together with heavy isotope searches, also rule out the mass range from a few GeV/c$^2$ to a few TeV/c$^2$ for dark matter SIMPS.

Although the above constraints have effectively closed off much of the interesting phase space for dark matter SIMPS, interest in them has not died out. Teplitz {\it et al.}\cite{tep} have recently proposed laboratory experiments for searching for anomalous heavy nuclei such as gold. Here, however, they specifically relax the constraint that SIMPS satisfy dark matter abundances. Also, SIMP motivations for dark matter have been revived in the form of strongly self-interacting particles which have been suggested as a means to circumvent astrophysical and cosmological problems perceived with the cold dark matter model\cite{sperg}.

\section{Searches for Fractional Charge, Stable Particles}
\noindent
In this section we ignore the restriction to massive particles. The discovery of a stable, fractional charge particle of any mass would be of great importance.
 
In the 1970's and 1980's there were four reviews of searches for fractional charge particles. General reviews, including accelerator and cosmic ray searches, were given by Jones\cite{jon} and Lyons\cite{lyo}. Smith\cite{smi2} primarily reviewed searches in terrestrial materials. Marinelli and Morpurgo\cite{mar} reviewed the magnetic levitation method. These reviews provide much information on search technologies. Our review is devoted to the results of the searches of the 1990's with older results included when still relevant.

We give all charges in units of the magnitude of the electron charge, $e$.

\subsection{Accelerator searches for fractional charge, stable particles}
\noindent
High energy, high intensity, electron-positron and proton-antiproton colliders are ideal instruments for searches for fractional charge particles. Surprisingly with respect to published results, full use has not been made of these opportunities. One limitation is that the experiments, or at least the published analyses, have concentrated on $q=\frac{1}{3}e,\mbox{ } \frac{2}{3}e,\mbox{ and }\frac{4}{3}e$. However in some cases the search may be more general.

\subsubsection{Searches using electron-positron colliders}
\noindent
The study of Z$^0$ decays provides a direct and an indirect way to search for fractional charge particle pair production at the Z$^0$:

\begin{equation}
\mbox{e}^+ + \mbox{e}^- \to \mbox{Z}^0 \to \mbox{X}^{+q} + \mbox{X}^{-q}.
\end{equation}

\noindent
However, there appear to be no published reports of {\it direct searches at the Z$^0$} for fractional charge particles produced according to Eq. 10.

The {\it indirect} search method is based upon the concept that if the X pair is {\it not} detected, then its decay width contributes to the invisible width of the Z$^0$. X particles with $q<1$ might not be detected because of their smaller than normal gas ionization in the track detectors. In units of neutrino width contribution, the direct measurement of the invisible width gives the quantity 3.07$\pm$0.12. Therefore, if fractional charge particles are pair produced but not detected at the Z$^0$, their weak interaction coupling must be smaller than that of the neutrino by a factor of order 2 or 3. This condition on the weak interaction coupling constant is thus a limit on the existence of a fractional charge particle with mass $\leq$ 45.5 GeV/c$^2$.
  
Two searches at the Z$^0$ using inclusive production have been reported. Buskulic {\it et al.}\cite{bus}, the ALEPH collaboration, searched 
for particles with $q=\frac{1}{3}e,\mbox{ } \frac{2}{3}e,\mbox{ or }\frac{4}{3}e$ in the mass range of 8 to 45 GeV/c$^2$. Their 90 \% C. L. upper limits are given in terms of $R=\sigma_{XX}/\sigma_{\mu\mu}$ where $\sigma_{\mu\mu}$ is the electroweak $\mu$ pair cross section. The upper limits on $R$ are in the range of 1 to 3 $\times$ 10$^{-3}$. Akers {\it et al.}\cite{ake}, the OPAL collaboration, reported similar null results.

Ackerstaff {\it et al.}\cite{ack}, the Opal collaboration, carried out a LEP search at higher energy, 130 to 183 GeV for pair produced X particles with $\frac{2}{3}e$. The X mass search range was 45 to about 90 GeV/c$^2$. The 95 \% C. L. upper limits on the cross section were in the range of 0.05 to 0.2 pb. 

There is an experimental uncertainty when searching at a collider for fractional charge particles that have quark-like charges. Is one actually searching for free quarks and, if so, what is the interaction cross section of the free quark? If it is a strongly interacting cross section then the particle may turn into conventional hadrons before its charge can be measured. Guryn {\it et al.}\cite{gur} considered this problem in their 1984 search, e$^+$ + e$^-$ $\to $ X$^{+q}$ + X$^{-q}$, at 29 GeV. They used a detector with a thickness of less than 1 \% of a conventional hadronic interaction length. No fractional charge particles were found , the search range extending to 13 GeV/c$^2$.

\subsubsection{Searches using proton-antiproton colliders}
\noindent
The highest energy reported search, p + $\bar{\mbox{p}} \to $ X$^{+q}$ + X$^{-q}$, for fractional charge particles is that of Abe {\it et al.}\cite{abe} carried out at the 1.8 TeV Tevatron. Figure 2 shows the 95 \% C. L. upper limits on the cross section for $q=\frac{2}{3}e \mbox{ and }\frac{4}{3}e$.

\subsubsection{Searches using heavy ion collisions}
\noindent
In the 1980's there were two accelerator searches using heavy ion collisions based on the concept that ``processes that involve heavy nuclei might yield fractional charge by mechanisms not possible in elementary particle interactions" to quote Lindgren {\it et al.}\cite{lin}. These experimenters examined the product of the collision of 1.9 GeV/nucleon $^{56}$Fe nuclei with Pb nuclei and gave an upper limit of less than 10$^{-4}$ fractional charge particles produced per Fe-Pb collision.

In another experiment, Barwick {\it et al.}\cite{bar2} examined the $14\leq Z \leq18$ projectile fragments of 1.8 GeV/nucleon $^{40}$Ar collisions searching for fractional charge particles with null results.

\subsection{Cosmic rays, halo particles, and searches for fractional charge, stable particles}
\noindent
The most comprehensive search to date for fractional charge particles impinging on the earth has been carried out by Ambrosio {\it et al.}\cite{amb2} using the MACRO experimental apparatus. They searched for lightly ionizing particles with $.25\leq{\beta}\leq 1.0$ and were sensitive to charges as small as $\frac{1}{5}e$. No fractional charge particles were found, the 90\% C.L. upper limit on the flux is of the order of

\begin{equation}
\phi\leq 2\times 10^{-14} \mbox{cm}^{-2}\mbox{sr}^{-1}\mbox{s}^{-1}.
\end{equation}
 
The limit as a function of $q$ is given in Fig. 7. Recall that the MACRO experiment lies beneath a mountain with a minimum overburden of 3300 m water equivalent. This leads to two limitations on the significance of these flux limits: strongly interacting fractional charge particles would not reach MACRO and non-strongly interacting particles must have sufficient energy to overcome the {\it dE/dx} loss.

Two other searches for fractional charge particles in cosmic rays and halo particles were carried out in the 1990's; both report upper limits on the flux only for particles with charges of $\frac{1}{3}e$ and $\frac{2}{3}e$.

Aglietta {\it et al.}\cite{agl} used the LSD detector in the Mont Blanc tunnel with an overburden of 5000 m of water equivalent to find the 90\% C.L. upper limits on the flux:

\begin{equation}
\phi(\frac{1}{3}e) \leq 2.3\times 10^{-13}\mbox{cm}^{-2}\mbox{sr}^{-1}\mbox{s}^{-1},\,\, \phi(\frac{2}{3}e) \leq 2.7\times 10^{-13}\mbox{cm}^{-2}\mbox{sr}^{-1}\mbox{s}^{-1}.
\end{equation}

Mori {\it et al.}\cite{mor} used the Kamiokande II detector with an overburden of 2700 m of water equivalent to find the 90\% C.L. upper limits on the flux:

\begin{equation}
\phi(\frac{1}{3}e) \leq 2.1\times 10^{-15}\mbox{cm}^{-2}\mbox{sr}^{-1}\mbox{s}^{-1},\,\, \phi(\frac{2}{3}e) \leq 2.3\times 10^{-15}\mbox{cm}^{-2}\mbox{sr}^{-1}\mbox{s}^{-1}.
\end{equation}

\noindent
Both sets of limits are shown in Fig. 7.

\begin{figure}[htb]
\begin{center}
\epsfig{file=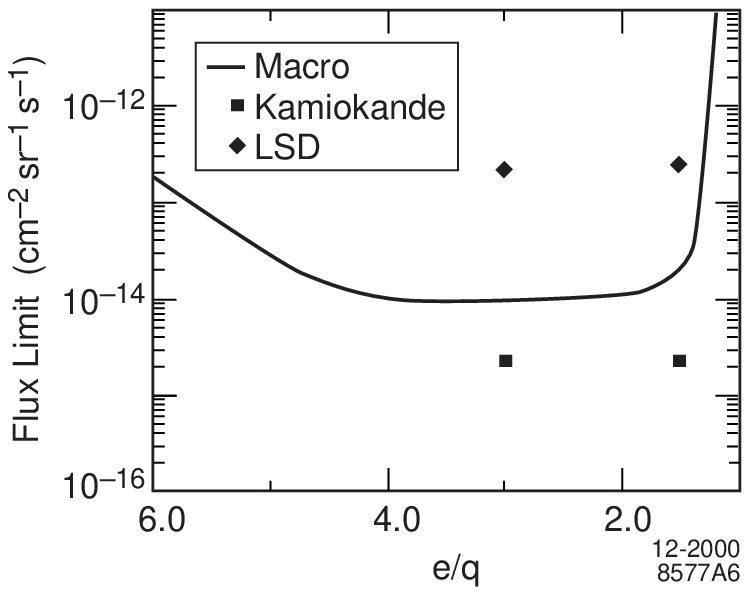,width=10cm}
\caption{The curve is the observed upper limit on the flux of fractionally charged particles as a function of $e/q$ from the MACRO detector, Ambrosio {\it et al.}\protect\cite{amb2}. The range of $\beta$ is 0.25 to 1.0. The points are upper limits from the Kamiokande II detector, Mori {\it et al.}\protect\cite{mor}, and the LSD detector, Aglietta {\it et al.}\protect\cite{agl}. $q$ is the fractional charge in units of the electron charge $e$. Thus $q=$1/2 correspond to $e/q=$2.}
\end{center}
\end{figure}

\subsection{Searches in ordinary matter for fractional charge, stable particles}
\noindent
At present there are two methods for searching for fractional charge, stable particles in ordinary matter- the levitometer method and the Millikan drop method. In the levitometer method, a small object consisting of ordinary matter is magnetically suspended in vacuum using either ferromagnetism\cite{smi2,mar} or superconductivity\cite{lar}. The object often, but not necessarily, is a sphere. The object's mass is in the range of 0.03 to 0.1 mg. An electric field is used to drive the object into forced oscillation and the charge on the object is measured by the resonant frequency of the oscillation.

The Millikan drop method goes back to his measurements of the electron charge\cite{mil}. In the modern development of the method\cite{joy,hend,perl1,nancy}, liquid drops with diameters in the range of 5 to 30 $\mu$m fall though air in the presence of an oscillating electric field. The drop charge is determined by measuring the drop's terminal velocity caused by the electric field.

Table 2 lists recent sensitive searches for fractional charge particles in ordinary matter. All experimenters reported null results except La Rue {\it et al.}\cite{lar} who reported finding $\frac{1}{3}e$ and $\frac{2}{3}e$ charges in niobium. This 1981 paper produced considerable interest in the possible existence of free quarks, but Smith {\it et al.}\cite{smi3}studied about four times as much niobium and found {\it no} evidence for any fractional charge particles in niobium. The La Rue {\it et al.}\cite{lar} report is not accepted today.  

\begin{table}
\begin{center}
\vspace*{12 pt}
\caption{Searches for fractional charge particles in ordinary matter. All experimenters reported null results except La Rue {\it et al.}\cite{lar}. See text. There are 6.4$\times10^{20}$ nucleons in a milligram. }
\vspace*{10 pt}
\begin{tabular}{|l|l|l|c|} \hline
\bf Method & \bf Experiment & \bf Material & \bf Sample Mass(mg)\\ \hline
\hline
superconducting levitometer & LaRue $et$ $al.$\cite{lar} & niobium & 1.1\\ \hline
ferromagnetic levitometer & Marinelli $et$ $al.$\cite{mar} & iron & 3.7\\ \hline
ferromagnetic levitometer & Smith $et$ $al.$\cite{smi3} & niobium & 4.9\\ \hline
ferromagnetic levitometer & Jones $et$ $al.$\cite{jon2} & meteorite & 2.8\\ \hline \hline
liquid drop & Joyce $et$ $al.$\cite{joy} & sea water & .05\\
\hline
liquid drop & Savage $et$ $al.$\cite{sav} & mercury & 2.0\\
\hline
liquid drop & Halyo $et$ $al.$\cite{hal} & silicone oil & 17.4\\
\hline
\hline
\end{tabular}
\vspace*{12 pt}
\end{center}
\end{table}

The sensitivity of the searches in Table 2 may be estimated by noting that there are 6.4$\times10^{20}$ nucleons in a milligram. The largest individual search sample is 17.1 mg of silicone oil, Halyo {\it et al.}\cite{hal}. They report that the concentration of particles with fractional charge more than 0.16$e$ from the nearest integer is less than 4.7$\times10^{-22}$ particles per nucleon with 95\% confidence.

As shown in Table 2, most materials used for fractional charge searches have been chosen for ease of use: niobium for superconducting levitation, iron for ferromagnetic levitation, mercury and oil for the Millikan drop method. These are probably not the best choices because fractional charge particles might easily be lost in electrically conducting materials or in refined materials such as oils. Lackner and Zweig\cite{lac1} have discussed the chemistry of atoms containing fractional charge and have discussed the most suitable materials for fractional charge searches\cite{lac2}. As summarized by Perl and Lee\cite{perl1} the most suitable materials appear to be meteoritic material from asteroids, terrestrial minerals that concentrate rare impurities and perhaps material from the moon's surface.

\subsection{ Searches for millicharge particles}
\noindent
All the fractional charge searches so far discussed are limited by their inability to detect very small charges, less than about $e$/10. This is because in all the methods the 3$\sigma$ or 4$\sigma$ limits on charge measurement precision are about $e$/10. A direct search for particles with charges in the range of 10$^{-1}e$ to 10$^{-6}e$ was carried out by Prinz {\it et al.}\cite{jaros} using a high energy electron beam to try to produce pairs of millicharge particles. The search, Fig. 8 from Prinz {\it et al.}\cite{jaros}, ruled out a mass region from 0.1 MeV/c$^2$ to 100 MeV/c$^2$, the 95\% C. L. upper limit on the charge being about 10$^{-5}e$ at the low mass boundary and 6$\times10^{-4}e$ at the high mass boundary. Figure 8 also shows other limits on the existence of millicharge particles\cite{dav1,golo,dav2}.

\begin{figure}[htb]
\begin{center}
\epsfig{file=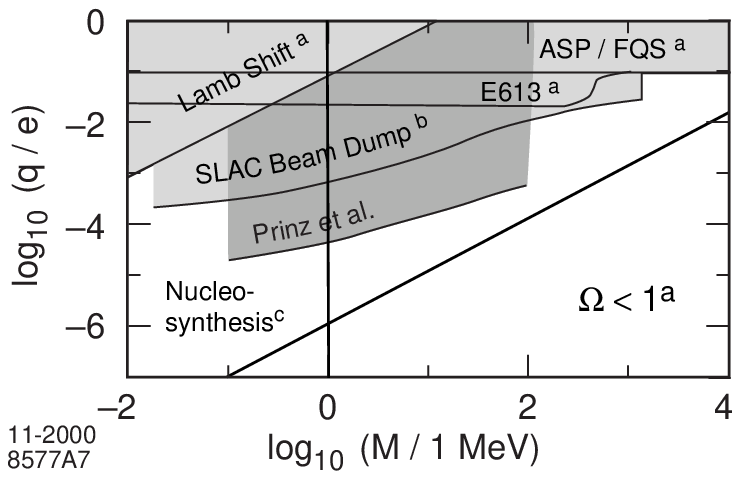,width=10cm}
\caption{Excluded sections of charge-mass parameter space\protect\cite{jaros}. The dark central region is the area excluded by Prinz {\it et al.}\protect\cite{jaros}. The lighter shaded regions give limits derived from other experiments. Other regions are excluded by astrophysical or cosmological arguments. The references are: (a) Davidson {\it et al.}\protect\cite{dav1}, (b) Golwich and Robinett\protect\cite{golo}, and (c) Davidson and Peskin\protect\cite{dav2}.}
\end{center}
\end{figure}

\section{Summary}
\noindent
There is no {\it confirmed} experimental or observational evidence for the existence of massive, stable, elementary particles other than the electron and proton. In particular, in the special case of dark matter there is no {\it confirmed} evidence for dark matter being composed of massive particles.

There are many areas where continued or new searches for massive particles are warranted:

\begin{itemize}
\item With respect to searches using proton-antiproton colliders, it seems to us that searches at the Tevatron are incomplete. Of course, thorough searches should be carried out when the Large Hadron Collider is operating.
\item With respect to searches using electron-positron colliders, it seems to us that there is uncertainty in the completeness of searches at the Z$^0$ peak and below the Z$^0$ peak. Of course, thorough searches should be carried out when an Electron-Positron Linear Collider is operating.
\item The various searches for fluxes of massive particles impinging on the earth, begun as dark matter explorations, should be continued even when the observed upper limit on the flux is too small to explain the expected dark matter density. There are two reasons. First, dark matter might consist of several types of massive particles. Second, the existence of a stable massive particle is of great importance even if it has nothing to do with dark matter.
\item Existing searches in bulk matter for massive particles with fractional or integer electric charge have probed to sensitivities in the range of 10$^{-22}$ to 10$^{-28}$ massive particles per nucleon. Improved search technology would allow substantial improvement of these sensitivities as well as the extension of the searches to matter from meteorites and the moon.
\end{itemize}


\noindent

\begin{thebibliography}{000}
\bibitem{gri}
K. Griest and M. Kamionkowski, Phys. Rev. Lett. {\bf 64}, 615 (1990).

\bibitem{kol}
E. W. Kolb {\it et al.},{\it Proc. Int. Conf. Dark Matter in Astro and Particle Phys.}, (Inst. of Phys., Bristol, 1999), Eds. H.V. Klapdor-Kleingrothanus and L. Baudis, p. 592.

\bibitem{der}
A. De R\'{u}jula {\it et al.}, Nucl. Phys. {\bf B333}, 173 (1990). 
 
\bibitem{dim}
S. Dimopoulos {\it et al.}, Phys. Rev. {\bf D41}, 2388 (1990).
 
\bibitem{jaros}
A. A. Prinz {\it et al.}, Phys. Rev. Lett. {\bf 81}, 1175 (1998).

\bibitem{pri}
J. R. Primack {\it et al.}, Ann. Rev. Nucl. Part. Sci. {\bf 38}, 751 (1988).

\bibitem{gai}
T.K. Gaisser, {\it Cosmic Rays and Particle Physics}, (Cambridge University Press, 1990).

\bibitem{cas}
C. Caso et al., Review of Particle Physics, The European Physical Journal C {bf 3}, 1 (1998).

\bibitem{bah}
N. Bahcall {\it et al.}, Science {\bf 284}, 1481 (1999).

\bibitem{l31}
M. Acciarri {\it et al.}, Phys. Lett. {\bf B462}, 354 (1999).

\bibitem{al1}
R. Borate {\it et al.}, Phys. Lett. {\bf B405}, 379 (1997).

\bibitem{de1}
P. Abreu {\it et al.}, Phys. Lett. {\bf B478}, 65 (2000).

\bibitem{abe}
F. Abe {\it et al.}, Phys. Rev. {\bf D46}, 1889 (1992).

\bibitem{bar}
S. W. Barwick {\it et al.}, Phys. Rev. Lett. {\bf 64}, 2859 (1990).

\bibitem{chu}
D.J.H. Chung {\it et al.}, Phys. Rev. Lett. {\bf 81}, 4048 (1998).

\bibitem{sno}
D. P. Snowden-Ifft {\it et al.}, Astrophys. J. {\bf 364}, L25 (1990).

\bibitem{ahl}
S. P. Ahlen {\it et al.}, MACRO Coll., Nucl. Instr. and Meth. {\bf A324}, 337 (1993).

\bibitem{mue}
G. P. Mueller, Phys. Lett. {\bf B245}, 681(1990).

\bibitem{amb}
M. Ambrosio et al., MACRO Coll., hep-ex/0009002.

\bibitem{fic}
D. J. Ficenec {\it et al.}, Phys. Rev. {\bf D36}, 311 (1987).

\bibitem{bari}
B. Barish {\it et al.}, Phys. Rev. {\bf D36}, 2642 (1987).

\bibitem{smi1}
P. F. Smith {\it et al.}, Nucl. Phys. {\bf B206}, 333 (1982).

\bibitem{hem1}
T. K. Hemmick {\it et al.}, Phys. Rev. {\bf D41}, 2074 (1990).

\bibitem{ver}
P. Verkerk {\it et al.}, Phys. Rev. Lett. {\bf 68}, 1116 (1992).

\bibitem{yam}
T. Yamagata {\it et al.}, Phys. Rev. {\bf D47}, 1231 (1993).

\bibitem{tur}
A. Turkevich {\it et al.}, Phys. Rev. {\bf D30}, 32 (1986).

\bibitem{dic}
W. J. Dick {\it et al.} , Phys. Rev. {\bf D33}, 1876 (1984).

\bibitem{chi}
R. S. Chivukula {\it et al.}, Phys. Rev. Lett. {\bf 65}, 957 (1990).

\bibitem{gou}
A. Gould {\it et al.}, Phys. Lett. {\bf B238}, 337 (1990).

\bibitem{jun}
G. Jungman {\it et al.}, Phys. Rept. {\bf 267}, 195 (1996).

\bibitem{ros}
L. Roszkowski, Nucl. Phys. Proc. Suppl. {\bf B87}, 21 (2000).

\bibitem{moral}
A. Morales, Nucl. Phys. Proc. Suppl. {\bf B87}, 477 (2000).

\bibitem{berg}
L. Bergstrom, Rept. Prog. Phys. {\bf 63}, 793 (2000).
 
\bibitem{bau}
L. Baudis and H.V. Klapdor-Kleingrothaus, {\it 2nd International Conference on Particle Physics}, (Institute of Physics, 2000), Eds. H.V. Klapdor-Kleingrothaus and I.V. Krivosheina, p. 881. 

\bibitem{ram}
Y. Ramachers, {\it 11th Recontres de Bois: Frontiers of Matter}, (Chateau de Bois, 1999), astro-ph/9911260.

\bibitem{pro}
{\it Proc.6th Int. Workshop on Topics in Astroparticle and Underground Physics}, (Paris, 1999), published in Nucl. Phys. Proc. Suppl. {\bf B87} (2000).

\bibitem{ell}
J. Ellis {\it et al.}, Phys. Rev. {\bf D62}, 075010 (2000).

\bibitem{abu}
R. Abusaidi {\it et al.}, Phys. Rev. Lett. {\bf 84}, 5699 (2000).

\bibitem{eva}
N. W. Evans {\it et al.}, astro-ph/0008156.

\bibitem{green}
A.M. Green, Phys. Rev. {\bf D63}, 043005 (2001).

\bibitem{ull}
P. Ullio and M. Kamionkowski, astro-ph/0006183.

\bibitem{copi}
C. J. Copi {\it et al.},, Phys. Lett. {\bf B461}, 43 (1999).

\bibitem{leh}
M.J. Lehner {\it et al.}, DRIFT coll., {\it 2nd Int. Conf. Dark Matter in Astro and Particle Physics, DARK98}, (Heidelberg, 1998), Ed. H.V. Klapdor-Kleingrothaus, p. 767.

\bibitem{gol}
S.R. Golwala {\it et al.}, Nucl. Instr. and Meth. Phys. Res. {\bf A444}, 345 (2000).

\bibitem{bern}
R. Bernabei {\it et al.}, Phys. Lett. {\bf B480}, 23 (2000) and references contained therein.

\bibitem{gerb} 
G. Gerbier {\it et al.}, Astropart. Phys. {\bf 11}, 287 (1999).

\bibitem{moral2}
A. Morales, astro-ph/9810341.

\bibitem{baud}
L. Baudis {\it et al.}, Phys. Rev. {\bf D63}, 022001 (2000).

\bibitem{gon}
R. Gondolo and J. Silk, Phys. Rev. Lett. {\bf 83}, 1719 (1999).

\bibitem{berg2}
L. Bergstrom, P. Ullio, J.H. Buckley, Astropart. Phys. {\bf 9}, 137 (1998). 

\bibitem{pros}
R.J. Protheroe, {\it Int. Sym. on High Energy Gamma-Ray Astronomy}, (Heidelberg, 2000), astro-ph/0011042.

\bibitem{ull2}
P. Ullio, astro-ph/9904086.

\bibitem{coutu}
S. Coutu {\it et al.}, Astropart. Phys. {\bf 11}, 429 (1999).
 
\bibitem{buff}
A. Buffington {\it et al.}, Astrophys. J. {\bf 248}, 1179 (1981).

\bibitem{super}
Super-Kamiokande Coll., {\it 30th International Conference on High-Energy Physics}, (Osaka, 2000), astro-ph/0007003.

\bibitem{good}
M.W. Goodman and E. Witten, Phys. Rev. {\bf D31}, 3059 (1985).

\bibitem{gold}
H. Goldberg and L. J. Hall, Phys. Lett. {\bf B174}, 151 (1986).

\bibitem{stark}
G.D. Starkman {\it et al.}, Phys. Rev. {\bf D41}, 3594 (1990).

\bibitem{mcg}
P.C. McGuire {\it et al.}, {\it Proc. Astrophysics Conf. in College Park, Maryland on Dark Matter}, (AIP, 1994), Eds. S. S. Holt and C.L. Bennett, p. 53.

\bibitem{bern2}
R. Bernabei {\it et al.}, Phys. Rev. Lett. {\bf 83}, 4918 (1999).

\bibitem{derb}
A.V. Derbin {\it et al.}, Physics of Atomic Nuclei {\bf 62}, 1886 (1999).

\bibitem{mohap}
R.N. Mohapatra and S. Nusinov, Phys. Rev. {\bf D57}, 1940 (1998).

\bibitem{tep}
V.L. Teplitz {\it et al.}, {\it Proc. 4th UCLA Symposium on Dark Matter DM2000}, (Marina del Rey, 2000), hep-ph/0005111.

\bibitem{sperg}
D. N. Spergel and P.J. Steinhardt, Phys. Rev. Lett. {\bf 84}, 3760 (2000).

\bibitem{jon}
L. W. Jones, Rev. Mod. Phys. {\bf 49}, 717 (1977).

\bibitem{lyo}
L. Lyons, Phys. Reports {\bf 129}, 225 (1985).

\bibitem{smi2}
P. F. Smith, Ann. Rev. Nucl. Part. Sci. {\bf 39}, 73 (1985).

\bibitem{mar}
M. Marinelli and G. Morpurgo, Phys. Reports {\bf 85}, 161 (1985).

\bibitem{bus}
D. Buskulic {\it et al.}, Phys. Lett. {\bf B303}, 198 (1993).

\bibitem{ake}
R. Akers {\it et al.}, Z. Phys.. {\bf C67}, 203 (1995). 

\bibitem{ack}
K. Ackerstaff {\it et al.}, Phys. Lett. {\bf B433}, 195 (1998).

\bibitem{gur}
W. Guryn {\it et al.}, Phys. Lett. {\bf B139}, 313 (1984).

\bibitem{lin}
M. A. Lindgren {\it et al.}, Phys. Rev. Lett. {\bf 51}, 1621 (1983).

\bibitem{bar2}
S. W. Barwick {\it et al.}, Phys. Rev. {\bf D30}, 691 (1984). 

\bibitem{amb2}
M. Ambrosio et al., MACRO Coll., Phys. Rev. {\bf D62}, 052003 (2000).

\bibitem{agl}
M. Aglietta {\it et al.}, Astroparticle Phys. {\bf 2}, 29 (1994).

\bibitem{mor}
M. Mori {\it et al.}, Phys. Rev. {\bf D43}, 2843 (1991).

\bibitem{lar}
G. S. LaRue {\it et al.}, Phys. Rev. Lett. {\bf 46}, 967 (1981).

\bibitem{mil}
R. A. Millikan, Phys. Rev. {\bf 32}, 349 (1911).

\bibitem{joy}
D. C. Joyce {\it et al.}, Phys. Rev. Lett. {\bf 51}, 1621 (1983).

\bibitem{hend}
C. D. Hendricks {\it et al.},Meas. Sci. Technol. {\bf 5}, 337 (1994).

\bibitem{perl1}
M. L. Perl and E. R. Lee, Am. J. Phys. {\bf 65}, 698 (1997).

\bibitem{nancy}
N. M. Mar {\it et al.}, Phys. Rev. {\bf D53}, 6017 (1996).

\bibitem{smi3}
P. F. Smith {\it et al.}, Phys. Lett. {\bf B153}, 188 (1985).

\bibitem{mari}
M. Marinelli and G. Morpurgo, Phys. Lett. {\bf B137}, 439 (1984).

\bibitem{jon2}
W. G. Jones {\it et al.}, Z. Phys. {\bf C43}, 349 (1989).

\bibitem{sav}
M. L. Savage {\it et al.}, Phys. Lett. {\bf B167}, 481 (1986).

\bibitem{hal}
V. Halyo {\it et al.}, Phys. Rev. Lett. {\bf 84}, 2576 (2000). 

\bibitem{lac1}
K. S. Lackner and G. Zweig, Phys. Rev. {\bf D28}, 1671 (1983).

\bibitem{lac2}
K. S. Lackner and G. Zweig, {\it Novel Results in Particle Physics}, (AIP Conf. Proc. 93, New York, 1982), Eds. R. S. Panvini, S. Alam, and S. E. Csorna.

\bibitem{dav1}
S. Davidson {\it et al.}, Phys. Rev. {\bf D43}, 2314 (1991).

\bibitem{golo}
E. Golowich and R. W. Robinett, Phys. Rev. {\bf D35}, 391 (1987).

\bibitem{dav2}
S. Davidson and M. Peskin, Phys. Rev. {\bf D49}, 2114 (1994).

\end{thebibliography}
\end{document}